\documentclass[]{article}

\pdfoutput=1 

\usepackage{amssymb}
\usepackage{amsmath,amsfonts}
\usepackage{algorithm}
\usepackage{algorithmic}
\usepackage{url}
\usepackage{cleveref}
\usepackage{eurosym}
\usepackage{todonotes}
\usepackage{xcolor}
\usepackage{pgfplots}
\pgfplotsset{compat = newest}
\usepackage{comment}
\usepackage{tikz}
\usepackage{authblk}
\usepackage[backend=bibtex]{biblatex} 
\title{Dynamic Rolling Horizon-Based Robust Energy Management for Microgrids Under Uncertainty}
\author[1]{Jens H\"onen}
\author[1]{Johann L. Hurink}
\author[2,3]{Bert Zwart}
\affil[1]{Faculty of EEMCS, University of Twente, Enschede, The Netherlands}
\affil[2]{Department of Mathematics and Computer Science, Eindhoven University of Technology, Eindhoven, The Netherlands}
\affil[3]{Centrum Wiskunde \& Informatica (CWI), Amsterdam, The Netherlands}
\date{}

\begin{document}

\maketitle

\begin{abstract}
Within the last few years, the trend towards more distributed, renewable energy sources has led to major changes and challenges in the electricity sector. To ensure a stable electricity distribution in this changing environment, we propose a robust energy management approach to deal with uncertainty occurring in microgrids. For this, we combine robust optimization with a rolling horizon framework to obtain an algorithm that is both, tractable and can deal with the considered uncertainty. The main contribution of this work lies within the development and testing of a dynamic scheduling tool, which identifies good starting time slots for the rolling horizon. Combining this scheduling tool with the rolling horizon framework results in a dynamic rolling horizon model, which better integrates uncertainty forecasts and realizations of uncertain parameters into the decision-making process. A case study reveals that the dynamic rolling horizon model outperforms the classical version by up to 57\% in costs and increases the local use of PV by up to 11\%.
\end{abstract}

\section{Introduction}
In the last decade, the energy transition has gotten more and more important, mainly due to climate change and its accompanying consequences. A central goal of this transition is the move towards renewable energy sources, such as e.g. wind and solar power. In combination with the increased penetration of electric vehicles (EVs) and heat pumps to replace fossil fuel-based mobility and heating, this transition has a large impact on current and future electricity systems. The increased and to some extent also synchronized loads of the new devices, as well as the uncontrollable and intermittent nature of renewable energy sources, create new severe problems, such as congestion in the low-voltage grid, or a mismatch between demand and supply.\\

To maintain a stable electricity distribution in the future, local energy management approaches, dealing with the new situation, are becoming increasingly popular, see, e.g. \cite{SurveyEnergyManagement2019Garciaetal}, \cite{SurveyMicrogridEMS2018Ziaetal}. Here, two aspects are important. The first is that these approaches allow to access different energy markets. The second is the choice of the considered time horizon. Energy markets allow access to the trading of energy and therefore are often used to represent the economic aspect of energy management approaches. Furthermore, different markets allow for different horizons of commitment regarding buying and selling decisions. It is essential that these decisions cannot be changed after submission. The second aspect concerns the choice of the considered time horizon. When only focusing on the next time slot of, e.g., 15 minutes, valuable information regarding future demand, supply, or market prices is omitted, leading to suboptimal or even infeasible solutions. The opposite approach, to consider longer time horizons (e.g., multiple days) at once, has the disadvantage that the considered forecasts and predictions for these longer horizons become increasingly inaccurate. Therefore, considering time-dependent uncertainty in data becomes a critical necessity to ensure feasibility.\\

Both aspects, the energy markets, and the time horizon, together with the involved uncertainty, lead to a highly dynamic planning problem, in which various decisions over different commitment horizons have to be made in an uncertain environment. The decisions can range from device management for batteries to trading decisions on a day-ahead or intraday market. In the case of an energy management approach for a residential neighborhood, also further aspects, such as privacy and communication have to be considered to ensure a high acceptance with participants.\\

Already for many years, robust optimization has gained attention in the area of energy management as a way to deal with uncertainty. The focus of robust optimization on the feasibility of the solution matches well with the risk-averse nature of current energy management systems. Therefore, robust optimization has successfully been applied to a wide range of settings, from classical unit commitment problems \cite{AROUnitCommitment2013Bertsimasetal}, \cite{ROUnitCommDynamicWind2017LorcaSun}, \cite{AROUnitCommitment2014XiongJirutitijaroen}, \cite{AROUnitCommitmentGenOutage2012XiongJirutitijaroen}, to energy management approaches within a microgrid \cite{AROisolatedMG2021Sadeketal}, \cite{AROuncertainArrivalEV2018ChoiHussainKim}, \cite{ROV2G2015BaiQiao} or energy trading schemes between microgrids \cite{ROMultiMicrogrid2016HussainBuiKim}, \cite{AROenergyTransactionsMultiMG2018Zhangetal}. The considered uncertainty varies from market prices \cite{RobustOptimalACDCPriceUncert2018HussainBuiKim}, to load and renewable energy sources \cite{AROuncertainArrivalEV2018ChoiHussainKim}, \cite{ROV2G2015BaiQiao}, \cite{ROMultiMicrogrid2016HussainBuiKim} or random failures in the grid structure \cite{AROUnitCommitmentGenOutage2012XiongJirutitijaroen}. The applied techniques can be split into two main groups, namely static and adaptive robust optimization. While in static robust optimization, all decisions have to be made directly at the start of the planning horizon, adaptive robust optimization allows some of the decisions to be postponed until a later point in time. This generally results in better solutions and also fits better with the dynamic nature of energy management problems.\\

However, one drawback of using robust optimization for longer time horizons with increased uncertainty over time, is that solutions tend to be very conservative. By combining (static) robust optimization with a rolling horizon framework, one can reduce this conservatism of solutions, while still taking future information into account. In general, the rolling horizon (or receding horizon) approach is a popular technique in academia and industry to solve (stochastic) large-scale optimization problems over a long time horizon, see e.g., \cite{RHSurvey2001ChandShuSethi}, \cite{RHTheory1991SethiSorger}. Hereby in all these optimization problems, the time horizon is discretized. Instead of solving the complete problem at once, a rolling horizon approach splits the time horizon into smaller, mostly overlapping time windows and solves the problem iteratively for these time windows. We refer to solving the problem for one such time window as an iteration of the rolling horizon approach. Usually, there are two main parameters characterizing the rolling horizon framework. The first one is the step size, which defines the distance between the starting times of two consecutive time windows and the second one is the length of the time windows. To cover the complete time horizon without gaps, it is obvious that the step size has to be smaller or equal than the time window length. Note, that in case the step size is smaller than the window length, only the partial solution up to the starting time of the next time window is realized.\\

Combining a rolling horizon framework with robust optimization ensures a feasible solution while reducing the considered uncertainty in data due to the shorter time horizons of the time windows. In addition, it allows for a natural integration of the realization of uncertain parameters into the decision-making process. This makes it possible to react to observed realizations of uncertainty even when using static robust optimization. Hence, the combination of static robust optimization and a rolling horizon framework yields similar properties compared to adaptive robust approaches while avoiding some of the additional computational complexity issues associated with adaptive robust techniques, see \cite{AROUncertainLinProg2004BenTaletal}, \cite{ROpracticalGuide2015GorissenYanikogludenHertog}. Another advantage of the combination of static robust optimization and a rolling horizon framework is the reduced communication between households and the microgrid, which is only needed when an iteration of the rolling horizon is started. This allows to limit the exchange of information to a fixed number of iterations per day.\\

This promising combination of a rolling horizon framework with a robust optimization approach has already been successfully used in energy systems applications, see e.g., \cite{ROUnitCommDynamicWind2017LorcaSun}, \cite{ARODynWind2015LorcaSun}, \cite{AROmobileEES2022Luetal}, \cite{ROMGensembeWind2017Craparoetal}, \cite{DROHEMSRH2020Wang}, \cite{ARORHisolatedMG2019Laraetal}, \cite{RHaaroSeasonalStorage2020Castellietal}. 
\begin{itemize}
\item In \cite{ROUnitCommDynamicWind2017LorcaSun} and \cite{ARODynWind2015LorcaSun}, dynamic uncertainty sets for wind forecasts, that are changing over time, are considered and by using either an affine dispatch policy, \cite{ROUnitCommDynamicWind2017LorcaSun}, or an alternating direction algorithm, \cite{ARODynWind2015LorcaSun}, the problem is solved within a rolling horizon framework.
\item In \cite{AROmobileEES2022Luetal} a mobile energy storage system and its decisions when and where to charge or discharge are modeled. Affine decision rules are used within a rolling horizon framework to deal with uncertainties in both, traffic as well as electricity demand and generation.
\item In \cite{ROMGensembeWind2017Craparoetal}, the impact of the length of the planning horizon of a scenario-robust microgrid energy management approach within a rolling horizon framework is investigated.
\item In contrast, in \cite{DROHEMSRH2020Wang}, also distributionally robust solution approaches for home energy management systems are proposed and solved in an iterative manner using a rolling horizon.
\item In \cite{ARORHisolatedMG2019Laraetal}, a two-stage model, combining a rolling horizon framework as the upper layer problem with an optimal power flow problem for the lower layer, is presented.
\item In \cite{RHaaroSeasonalStorage2020Castellietal}, the day-ahead unit commitment and economic dispatch problem of a multi-energy system is modeled and solved using a combination of adaptive robust optimization and a rolling horizon framework.
\end{itemize}

One key similarity between the above-mentioned approaches is the static repetition of iterations of the rolling horizon based on a fixed step size. This step size directly translates into a fixed number of daily iterations, which in our setting may be bounded by the limited communication between households and the microgrid. However, given these limited iterations, it is important that each individual iteration should create as much value to the overall solution as possible to justify this information exchange.\\

Our work directly aims at this critical aspect. Instead of controlling the re-optimization of the energy management problem simply based on the step size, we extend the rolling horizon framework by a dynamic scheduling component. This component allows us to better integrate structural knowledge of the considered time-dependent uncertainty in the case of a limited number of daily iterations into the decision-making process. In addition, this extended rolling horizon framework may also be a relevant concept to be applied to other stochastic problems with a limited number of re-optimizations due to some scarce resources. \\

The key contributions of this work are:
\begin{enumerate}
\item We provide a detailed analysis of the impact of uncertainty on the performance of a classical rolling horizon within the context of a dynamic energy management problem. The analysis reveals that in particular, the time-dependent nature of the photovoltaic (PV) uncertainty as well as the uncertain EV demand is responsible for the majority of the improvements of the rolling horizon solutions over a fully static solution.
\item Based on these results, we analyze the structure of the time-dependent uncertainty sets of the PV generation, as well as the  EV uncertainty, which highlights the importance of certain time slots, while others may be discarded without loss of quality in the solution.
\item Based on these insights, we develop a dynamic scheduling tool to identify promising starting time slots for the iterations of a generalized rolling horizon. This scheduling tool is based on a version of the well-known knapsack problem from combinatorial optimization, in which items correspond to starting time slots of the rolling horizon. The combination of the dynamic scheduling tool with the main principles of the rolling horizon framework leads to a new, dynamic rolling horizon approach, which allows a more flexible scheduling of its iterations. 
\item A detailed numerical simulation, including discussion and comparison of the dynamic and classical rolling horizon, is carried out to show the performance improvements of the scheduling tool.
\end{enumerate}

This work is an extension of earlier work on a local robust energy management approach (see \cite{RobustEMMG2022Hoenenetal}) and is therefore based on a similar setting and microgrid model. We extend the model by further details, such as, e.g., charging and discharging efficiencies or different time slot lengths for the markets. The main additional difference lies in the considered techniques from robust optimization, as well as the usage of a rolling horizon.\\ 

This paper is structured as follows: In \Cref{SSetting}, we introduce the setting and explain the properties of the considered markets, devices, and participants. Next, a mathematical formulation of the problem is given in \Cref{SModel}, followed by a detailed analysis of the classical rolling horizon model in \Cref{SResultsCRH}. In \Cref{SResultsDRH}, we propose the new dynamic scheduling tool and compare the classical and dynamic rolling horizon models with each other. We conclude the work in \Cref{SConclusion} by providing future research directions and open questions.

\section{Setting}\label{SSetting}
The setting considered in this paper can be described as an energy planning problem for a residential microgrid with access to two different electricity markets for a discretized time horizon $\mathcal{T}$ of up to one week. The planning problem is defined from the perspective of the \textit{microgrid operator} (MGO), who is responsible for fulfilling the electricity demand of all members of the microgrid. To ensure this, for each time slot of the considered time horizon, the electricity supply has to be at least as large as the occurring demand. Hereby, the MGO acts at the electricity markets as a representative of the microgrid to buy and sell electricity, as well as to manage certain devices within the microgrid.\\

\begin{figure}[ht]
\centering
\includegraphics[width=8.5cm]{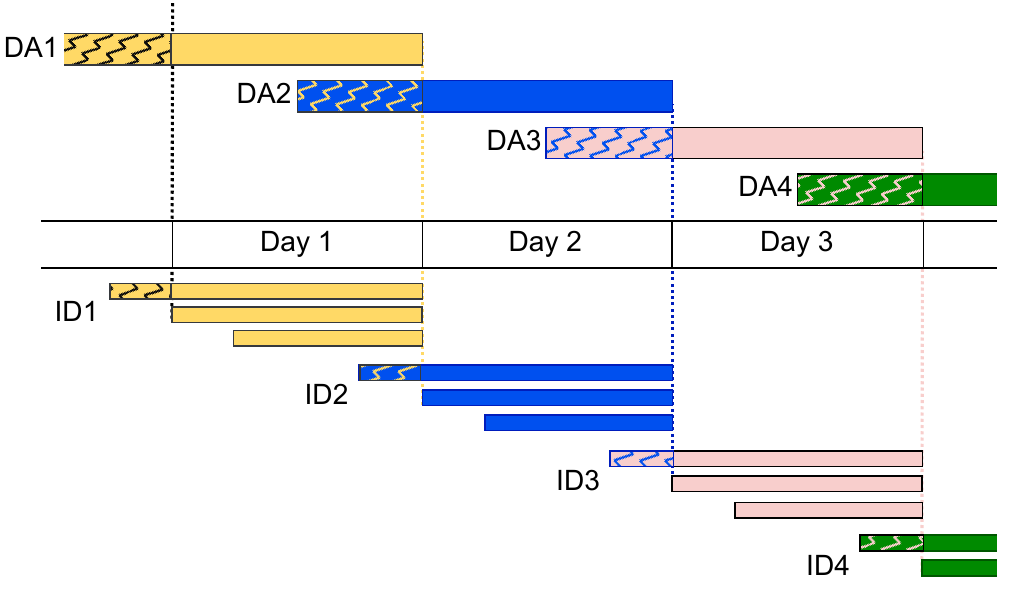}
\caption{Market Decisions: The upper level displays the iterations of the rolling horizon, which is responsible for the decisions of the day-ahead market. The first part of each rolling horizon iteration represents the already fixed day-ahead market decisions of the previous iteration. The bottom part displays the folding horizon iterations and its decisions regarding the intraday market and devices.}
\label{Fig1Market}	
\end{figure}

For the two markets, we consider the \textit{day-ahead market} and an \textit{intraday market}, similar to current electricity systems. As the names already indicate, decisions within the day-ahead market have to be made on the day before the actual delivery. The electricity can be bought or sold for one-hour time slots for a given price of $\pi^{DA}_t$, whereby within each hour, the amount is assumed to be evenly distributed. Given the time distance between making the market decisions and acting accordingly to them, deviations may occur, such as e.g., a change in the actual demand of the microgrid for a specific time. To be able to react to these changes, we assume that on the intraday market, electricity for time $t$ can be bought and sold till the end of the previous time slot $t-1$. In contrast to the day-ahead market, the intraday market operates on 15-minute time slots, providing an opportunity to react to smaller changes. Unless explicitly specified, we refer to the 15-minute intervals as time slots. Buying and selling prices on the intraday market are denoted with $\pi^{ID,b}_t$, respectively $\pi^{ID,s}_t$. Similarly to the intraday market, device decisions can also be made at the start of the corresponding time slot. \Cref{Fig1Market} indicates at which times the decisions for the different markets have to be made and for how long these decisions are in place. \\

The rolling horizon integration is strongly orientated on the decisions of the day-ahead market. Due to the requirements and regulations of this market, an iteration starts at 12 pm every day to submit the day-ahead decisions for the following day, see \Cref{Fig1Market}. The corresponding time window has a length of 36 hours to cover the complete next day for the day-ahead market decisions. In between the day-ahead market iterations of the high-level rolling horizon, a folding horizon approach is used to adapt the short-term decisions of the intraday market and the considered devices to observed realizations and updated forecasts of the uncertainty. Within the folding horizon, a step size $t_0$, ranging from 15 minutes to 1 day, is used to schedule the folding iterations. Note that the end of the time window is equal to the end of the last submitted day-ahead market decisions, leading to time window lengths between 12 and 36 hours, see also \Cref{Fig1Market}. Within the folding iterations, the day-ahead market variables are already fixed to the previously submitted decisions. Within the following, we do not explicitly differentiate between iterations of the rolling and folding horizon. The pseudo-code of the rolling horizon implementation can be seen in \Cref{Alg1}.\\

\begin{algorithm}
\caption{Classical Rolling Horizon Approach}
\label{Alg1}
\begin{algorithmic}
\FOR{$t=1$ to $T$ with step size $t_0$}
\STATE solve $staticRobustModel(t,data)$
\STATE reveal uncertainty realizations in $\left[ t,t+t_0\right) $
\STATE compute and fix new initial values in $data$
\IF{$t$ starts at 12:00}
\STATE fix day-ahead market decisions for the following iteration
\ENDIF
\STATE save the decisions in $\left[ t,t+t_0\right) $ to \textit{final decisions}
\ENDFOR
\STATE compute the actual costs based on the realizations and the decisions from \textit{final decisions}
\RETURN costs and \textit{final decisions}
\end{algorithmic}
\end{algorithm}

To model bottlenecks in the infrastructure, as well as possible restrictions by the Distribution System Operator (DSO), we assume that there is a capacity limit $C^{grid}_t$ for the line connecting the microgrid to the main grid. In addition to the decisions regarding the markets, the MGO is also able to use the flexibility offered by devices within the microgrid.\\ 

Within this work, we focus on a few selected types of devices, which either offer flexibility or have a large impact on the system, and are therefore key components in any future energy management system. W.l.o.g., let $\mathcal{N}_x$ denote the set of device type $x$:
\begin{itemize}
\item Prosumer Load: Let $p^{L,i}_t$ denote the electricity usage of prosumer $i \in \mathcal{N}_P$ during time slot $t$. We assume that this load is inflexible, i.e. it cannot be changed. Examples of such load sources are lighting, cooking, or TV. Other electricity sources, such as a dishwasher or the washing machine, which are in principle able to shift their load in time, by starting earlier or later, are within the context of this work also considered to be inflexible. 
\item PV system: Let $p^{PV,j}_t$ be the electricity, which PV system $j\in \mathcal{N}_{PV}$ produces at time $t$. We assume that this production is curtailable, meaning that only an amount $0 \leq x^{PV,j}_t \leq p^{PV,j}_t$ is actually used.
\item Battery: A battery $k\in \mathcal{N}_B$ is a device that can be used to shift the demand or production of electricity through time. It is specified by several parameters, namely the capacity $C^{B,k} \geq 0$, the charging and discharging limits $CL^{B,k}, DL^{B,k} \geq 0$, as well as the initial state of charge $SOC^{B,k}$. For technical reasons, we assume that the target state of charge of any battery at the end of the planning horizon is equal to its initial state of charge. In addition to this, we consider losses due to charging and discharging, represented by the charging and discharging efficiency $CE^{B,k}$ and $DE^{B,k}$.
\item EV: An EV $h\in \mathcal{N}_{EV}$ can be considered as a battery with additional limitations. As before, we have a capacity $C^{EV,h}$, a charging and discharging limit $CL^{EV,h}$, $DL^{EV,h}$, charging and discharging efficiencies $CE^{EV,h}, DE^{EV,h}$ and an initial state of charge $SOC^{EV,h}$. In contrast to a battery, an EV may not always be available. Let the times, in which EV $h$ cannot be charged or discharged be given by $\mathcal{I}(h) \subseteq \mathcal{T}$. In addition, we have a given demand $p^{EV,h}_t\geq 0$ for time slot $t$. In a time slot where the EV arrives home, this value specifies the used energy during this trip, and in all other cases, it is 0.
\end{itemize}

Due to the different time scales, for which decisions have to be made on the two markets and on the device level, as well as the intermittent nature of renewable energy sources and human behavior, some of the parameters within this planning problem are difficult to predict and deviations from predicted values may appear. In the following, we present the different sources of uncertainty we consider within this work:
\begin{itemize}
\item Load: The household load strongly depends on the behavior and actions of the residents, which can be subject to sudden changes, which are not perfectly predictable, see e.g., \cite{LoadUncertaintyForecastLSTM2020Alhusseinetal}.
\item PV generation: Even though (PV) forecasting methods are getting better, currently it is not possible to perfectly predict the PV generation for multiple days in advance, see e.g., \cite{PVforecasting2019Hanetal}. Hence, PV generation is subject to uncertainty, whereby short-term forecasting often is more accurate than long-term forecasting of 24 hours and more. 
\item EV demand: Even though the electricity demand for driving a certain distance can be computed quite accurately, in practice various aspects, such as the outside temperature, vehicle heating or cooling, or the traffic, have an influence on the actual demand, see e.g., \cite{EVdemandforecasting2020Ghotgeetal}, \cite{EVdemandforecasting2022Wuetal}. 
\item EV arrival and departure times: In settings with short time slots, such as e.g. 15 minutes, already smaller deviations due to crowded roads may lead to uncertainty in arrival or departure times. In addition, unplanned events like working longer, or a small detour to a supermarket can be other sources of uncertainty.
\item Market prices: The prices of both considered markets are dependent on demand and supply, which are both not perfectly predictable, leading to fluctuations in the prices, see e.g., \cite{DAPriceforecasting2021LiBecker}, \cite{IDPriceforecasting2019Uniejewskietal}, \cite{Priceforecasting2020Jahangiretal}. 
\end{itemize}

\section{Mathematical Formulation}\label{SModel}
In the first step, we introduce the deterministic base model, and based on this in the second step, we present how uncertainty sets can be modeled and integrated.

\subsection{Deterministic Model}\label{SSDetModel}
In the following, we first define the variables and then describe the microgrid model, consisting of the objective function (\ref{M1OF}), as well as the constraints (\ref{M1BalanceC}) -- (\ref{M1MC2}). Due to the similarity in the model given in \cite{RobustEMMG2022Hoenenetal}, we follow the notation used there.
\subsubsection{Variables}
\begin{itemize}
\item $x^{DA,buy}_t\geq 0$, $x^{DA,sell}_t \geq 0$: total amount of electricity bought or sold at the day-ahead market for time slot $t$.
\item $x^{ID,buy}_t \geq 0$, $x^{ID,sell}_t \geq0$: total amount of electricity bought or sold at the intraday market for time slot $t$.
\item $x^{PV,j}_t\geq0$: amount of electricity from PV system $j$, which is used during time slot $t$.
\item $x^{B,k}_{C,t} \geq 0$, $x^{B,k}_{D,t} \geq 0$: amount of charged and discharged electricity of battery $k$ during time slot $t$.
\item $x^{EV,h}_{C,t} \geq 0$, $x^{EV,h}_{D,t} \geq 0$: amount of charged and discharged electricity of EV $h$ during time slot $t$.
\end{itemize}

\subsubsection{Objective Function}
The objective function (\ref{M1OF}) represents the total cost (in \euro) of the microgrid. It can be split up into costs for the day-ahead market and the intraday market.
\begin{align}\label{M1OF}
\text{min} & \sum_{t \in \mathcal{T}} \bigg( \pi^{DA}_t \left( x^{DA,b}_t -x^{DA,s}_t \right)  + \pi^{ID,b}_t x^{ID,b}_t - \pi^{ID,s}_t x^{ID,s}_t \bigg)
\end{align}

\subsubsection{Demand-Supply-Balancing Constraint}
Constraint (\ref{M1BalanceC}) ensures that the total supply matches the overall demand for all time slots $t \in \mathcal{T}$.
\begin{align}\label{M1BalanceC}
& \sum_{j \in \mathcal{N}_{PV}}x^{PV,j}_t - \sum_{k\in \mathcal{N}_B} \left( x^{B,k}_{C,t} - x^{B,k}_{D,t} \right) - \sum_{h \in \mathcal{N}_{EV}}\left( x^{EV,h}_{C,t} - x^{EV,h}_{D,t} \right) + \nonumber \\
&   y^{DA,b}_t + x^{ID,b}_t - y^{DA,s}_t - x^{ID,s}_t = \sum_{i \in \mathcal{N}_P} p^{L,i}_t 
\end{align}

\subsubsection{PV Constraint}
The PV constraint (\ref{M1PVC1}) ensures that the amount of electricity used from the PV does not exceed the PV production.
\begin{equation}\label{M1PVC1}
0 \leq x^{PV,j}_t \leq p^{PV,j}_t \quad \forall t \in \mathcal{T}, j \in \mathcal{N}_{PV}
\end{equation}

\subsubsection{Battery Constraints}
Constraint (\ref{M1BC1}) ensures that the state of charge (SoC) of battery $k \in \mathcal{N}_B$ is always between 0 and its total capacity. It includes the losses due to charging and discharging. Constraints (\ref{M1BC3}) and (\ref{M1BC4}) limit the charging and discharging for each time slot $t$. This model is a slightly adapted version of the well-known SoC book-keeping model, see e.g., \cite{ARORHisolatedMG2019Laraetal}.
\begin{align}
0 \leq SOC^{B,k} +  \sum_{s=1}^{t} \Bigg( CE^{B,k}x^{B,k}_{C,s} - \dfrac{x^{B,k}_{D,s}}{DE^{B,k}} \Bigg)  & \leq C^{B,k}  \label{M1BC1}\\
0 \leq x^{B,k}_{C,t} &  \leq CL^{B,k}  \label{M1BC3}\\
0 \leq x^{B,k}_{D,t} &  \leq DL^{B,k}  \label{M1BC4}
\end{align}

\subsubsection{EV Constraints}
Due to the similarities between an EV $h \in \mathcal{N}_{EV}$ and a battery, the constraints are rather similar to each other. The additional constraint (\ref{M1EVC5}) ensures that an EV $h \in \mathcal{N}_{EV}$ can only be charged or discharged when it is connected to the grid. 
\begin{align}
0  \leq SOC^{EV,h} + \sum_{s=1}^{t} X^{EV}_t  & \leq C^{EV,h}  && \forall t \in \mathcal{T} \label{M1EVC1}\\
0 \leq x^{EV,k}_{C,t} &  \leq CL^{EV,k} && \forall t \in \mathcal{T}  \label{M1EVC3}\\
0 \leq x^{EV,k}_{D,t} &  \leq DL^{EV,k} && \forall t \in \mathcal{T}  \label{M1EVC4} \\
x^{EV,h}_t & = 0 &&\forall t \in \mathcal{I}(h) \label{M1EVC5} ,
\end{align}
whereby $X^{EV}_t = CE^{EV,h}x^{EV,h}_{C,s} - \dfrac{x^{EV,h}_{D,s}}{DE^{EV,h}} - p^{EV,h}_s$.\\

\subsubsection{Market and Grid Constraints}
Constraints (\ref{M1MC1}) and (\ref{M1MC2}) ensure that the sum of bought or sold electricity does not exceed the line capacity connecting the microgrid with the markets.
\begin{align}
0 \leq x^{DA,buy}_t + x^{ID,buy}_t &\leq C^{grid} && \forall t \in \mathcal{T} \label{M1MC1}\\
0 \leq x^{DA,sell}_t + x^{ID,sell}_t &\leq C^{grid} &&  \forall t \in \mathcal{T} \label{M1MC2}
\end{align}

\subsection{Robust Model}
To integrate the mentioned uncertainties into the model presented in \Cref{SSDetModel}, we first introduce the types of used uncertainty sets for the different uncertain parameters. We start with standard uncertainty sets, which make up the core of any robust optimization approach. There are a couple of well-studied uncertainty sets, for which tractable robust counterparts have been established. See \cite{ROpracticalGuide2015GorissenYanikogludenHertog} and \cite{ROTheoryApplication2011Bertsimas} for further information and a detailed introduction to static robust optimization. For our model, we make use of two uncertainty sets, namely the budget and the box uncertainty set:
\begin{itemize}
\item Box uncertainty set: The EV demand is an example of a box uncertainty set. Assume that for an EV $h$, the demand $p^{EV,h}_t>0$ in time slot $t$ is positive. Then due to various reasons, there may be a (small) deviation of the used energy from the expected value $p^{EV,h}_t$. Rather than defining an uncertainty interval for all possible values of this parameter, we first split the EV demand up into a certain, known part and an uncertain part
\begin{displaymath}
p^{EV,h}_t = \hat{p}^{EV,h}_t(1+\alpha^{EV,h}_t u^{EV,h}_t),
\end{displaymath}
where $\hat{p}^{EV,h}_t$ corresponds to the expected or predicted part and $\alpha^{EV,h}_t u^{EV,h}_t$ models the uncertain part, with $u^{EV,h}_t \in \left[ -1,1\right]$ being a random variable to model the true realization. Furthermore, $\alpha^{EV,h}_t$ defines (together with $\hat{p}^{EV,h}_t$) how large the uncertainty interval is. Hence, the actual uncertainty is now completely covered by the variable $u$, and the uncertainty set for the EV demand of a given EV $h$ is given by
\begin{displaymath}
\mathcal{U}^{EV,h} = \left\lbrace u \in \mathbb{R}^{\vert \mathcal{T} \vert} \vert \Vert u \Vert_{\infty} \leq 1 \right\rbrace .
\end{displaymath}
\item Budget uncertainty set: The household load is an example of a budget uncertainty set. In addition to the box uncertainty set, the budget uncertainty set adds another constraint, which limits the number of realizations at extreme points (of the box uncertainty set). Similar to the EV demand, we first modify the uncertain parameter to
\begin{displaymath}
p^{L,i}_t = \hat{p}^{L,i}_t(1+\alpha^{L,i}_t u^{L,i}_t).
\end{displaymath}
Hereby the box uncertainty part of the load is modeled in the same manner as for the EV demand. We further restrict the box uncertainty set of all prosumers for time slot $t$ by adding the following constraint
\begin{displaymath}
\sum_{i \in \mathcal{N}_{EV}}\vert u^{L,i}_t \vert \leq \Gamma^{L}_t.
\end{displaymath}
This constraint ensures that at most $\Gamma^L_t$ prosumers are at their own respective extreme points ($ u^{L,i}_t=1$ or $ u^{L,i}_t=-1$) w.r.t. their load. Combining the constraints, we get the following budget uncertainty set
\begin{displaymath}
\mathcal{U}^{L}_t = \left\lbrace u \in \mathbb{R}^{\vert \mathcal{N}_P \vert} \vert \Vert u \Vert_{\infty} \leq 1, \Vert u \Vert_{1} \leq \Gamma^L_t \right\rbrace.
\end{displaymath}
\end{itemize}

The remaining uncertainty sets can be modeled similarly, with the PV generation being a budget uncertainty set, and the market prices and the EV arrival and departure times being box uncertainty sets.\\

In the following, we shortly explain how to include the uncertainty sets into the constraints and then reformulate these constraints to get tractable formulations. For this, we use the demand-supply-balance constraint (\ref{M1BalanceC}) for some fixed time slot $t$ as an example. W.l.o.g., we replace the left-hand side with $X$ for the sake of readability. Then, the deterministic constraint (\ref{M1BalanceC}) is given by:
\begin{align*}
X = \sum_{i \in \mathcal{N}_P} p^{L,i}_t .
\end{align*}
Replacing the load with its uncertain version yields
\begin{align*}
X = \sum_{i \in \mathcal{N}_P} \hat{p}^{L,i}_t \left( 1+\alpha^{L,i}_t u^{L,i}_t\right) \quad \forall u \in \mathcal{U}^L_t.
\end{align*}
This would instantly lead to a problem formulation with an empty set of feasible solutions, as $X$ has to be equal to multiple different values, assuming that $\alpha \neq 0$ and $p^L\neq 0$. To avoid such a situation, we first replace the equality sign with a $\geq$ and indirectly assume that the overproduction of electricity does not lead to problems. A second problem is that we are not able to explicitly write down the above constraints for all $u$, as $\mathcal{U}^L_t$ contains infinitely many points $u$. Hence, we need a technique to reformulate the constraints in a tractable manner. This reformulation is called a robust counterpart, and for many different uncertainty sets, this process is well-known and understood. For the budget uncertainty set, the resulting robust counterpart is given by
\begin{align*}
X \geq \sum_{i \in \mathcal{N}_P} \hat{p}^{L,i}_t + \Vert y \Vert_{1} + \Gamma^L_t \Vert \hat{p}^L_t \alpha^L_t - y \Vert_{\infty},
\end{align*}
where $y$ is a variable vector of length $\vert \mathcal{N}_P \vert$, and $\hat{p}^L_t \alpha^L_t$ is an element-wise multiplication of the two vectors. Both, the $L1$ and the $L$-infinity norm can be reformulated using linear constraints and auxiliary variables. Hence, the resulting formulation is a slightly larger linear program. Following the same procedure for the remaining constraints, as well as the objective function, we end up with the robust counterpart of the complete model. For further details and examples regarding robust optimization, see \cite{ROMethodologyApplications2002BenTalNemirovski}.

\section{Results of the Classical Rolling Horizon}\label{SResultsCRH}
In this section, we test and analyze the proposed classical rolling horizon model for several scenarios and compare it to a fully static robust model, which considers the complete time horizon at once. We analyze the differences in detail and present insights into the working of the rolling horizon framework in the presence of uncertainty. Throughout this first analysis, we mainly focus on the costs of the different solutions, as these are making up the objective function of the models.

\subsection{Simulation}\label{SSSimulation}
All following numerical experiments are based on the setting described in \Cref{SSetting}. The considered time horizon is 3 days, discretized into 15-min time slots. The standard test case is a microgrid of 20 households with in total 15 EVs, 17 PV systems, and one communal battery system.

\subsubsection{Data}
The household load is modeled based on the average Dutch household load from the 12th to the 14th of April 2021. The data is publicly available \cite{NEDUDemandProfiles} and can be adapted to the total household demand of one year. For this study, we considered an average electricity usage of 3,500 kWh per year and household, which yields a consumption of 8 to 10kWh per day and household. The PV systems are modeled such that in the best case the daily production is around 11kWh. The predicted PV profile follows the curve of a sunny day without any clouds, although slightly lower, to allow for a better integration of the uncertainty set. The EVs are modeled based on the VW ID.3, with a battery capacity of 58 kWh and a charging and discharging limit of 11 kW, or 2.75 kWh per 15-min time slot. Regarding the electricity demand of the EV, we assume a distance of 20 to 70 km per trip and an electricity demand of 18kWh/100km. The communal battery system is modeled after 3 connected Telsa Powerwall batteries with an aggregated capacity of 42 kWh and charging and discharging limits of 15 kW or 3.75 kWh per time slot, \cite{TeslaPowerWall}. Charging and discharging efficiencies are similar to the EVs with both being $95\%$, resulting in a round trip efficiency of about $90\%$. The battery system is scaled such that the discharging power can handle the maximal aggregated household load per time slot. The initial state of charge is assumed to be 0, and this is also the required state of charge at the end of the time horizon. The prices for the day-ahead market from the 12th to the 14th of April 2021 have been accessed from \cite{DAMarketPrices}, while the Dutch intraday market prices were obtained from the Dutch TSO TenneT \cite{IDMarketPrices}.\\

\subsubsection{Uncertainty}
We introduce 3 different uncertainty scenarios $A$, $B$, and $C$ with scenario $B$ being the standard one, $A$ having less uncertainty, and $C$ more (see \Cref{TabScenarios} for the concrete parameter values).\\

\begin{table}[ht]
\centering
\caption{Uncertainty Scenarios}
\label{TabScenarios}
\begin{tabular}{c | c c c c c} 
Scenario & $\alpha^L$ & $\alpha^{PV}$ & $\alpha^{EV}$ & $\alpha^{DA}$ & $\alpha^{ID}$ \\ 
\hline
$A$ & 0.10 & 0.10 & 0.05 & 0.10 & 0.20 \\ 
$B$ & 0.20 & 0.25 & 0.10 & 0.15 & 0.35 \\
$C$ & 0.35 & 0.40 & 0.20 & 0.20 & 0.50 \\
\end{tabular}
\end{table} 

While for many uncertain parameters, the uncertainty sets are built up as static sets (see \Cref{SModel}), there are also parameters, for which the forecasts may change over time. For such parameters, the static uncertainty sets need to be updated with improved forecasts to reflect the dynamic nature of uncertainty. Within this work, the PV generation is one example of such an uncertain parameter. It has been shown that long-term weather forecasts are less accurate than short-term ones. This difference can be attributed to different forecasting techniques, see \cite{SurveyPVForecastTechOpt2020Ahmedetal}, \cite{SurveyPVForecast2021SinglaDuhanSaroha}, \cite{PVforecasting2020Ahmedetal} among others. Within this work, we assume that the next two hours can be predicted with improved accuracy, while the remaining forecasts follow the long-term forecast with lower accuracy. This translates into larger uncertain sets for time slots of two or more hours into the future, and smaller uncertainty sets for the next two hours. Within these two hours, the uncertainty sets get smaller the closer they are to the time of the predicted parameter.\\

Within the following simulations, all uncertainty realizations are drawn i.i.d. from the uniform distribution $\mathcal{U}(-1,1)$. The realizations of a given time slot $t$ are only revealed when the iterations of the rolling horizon model are already beyond this point in time to simulate a setting, in which only past realizations can be observed.

\subsection{Results and Discussion}\label{SSBaseComparison}

Starting with base scenario $B$, we first compare the objective values for various step sizes of the rolling horizon model with the fully static model, which solves the complete time horizon of three days at once. The tested step sizes in time slots are $\vert \mathcal{T}\vert$, 96, 48, 24, 16, 12, 8, 4, and 2. Note that the step size $\vert\mathcal{T}\vert$ solves the fully static model at once, while the step size of 96 implies that the iterations of the rolling horizon model are equivalent to the decision-making time slots of the day-ahead market, that is at 0 and 12 pm on day 1, and 12 pm on day 2 (see also \Cref{Fig1Market} from \Cref{SSetting}). To account for the influence of random realizations of the uncertainty, we run the models 5 times and focus on the averaged values, as shown in \Cref{Tab1BaseComparisonFull}.\\

\begin{table}[h]
\centering
\caption{Comparison of step sizes, the fully static model, Scenario $B$}
\label{Tab1BaseComparisonFull}
\begin{tabular}{c | c c c} 
Step size & avg. obj. value & abs. improv. & rel. improv. \\
\hline
$\vert \mathcal{T}\vert$ & 4.6367 & 0.000 & 0.0\%  \\ 
96 & 4.9705 & -0.3338 & -7.2\%  \\
48 & 4.3666 & 0.2701 & 5.8\% \\
24 & 4.2409 & 0.3958 & 8.5\%  \\ 
16 & 3.9931 & 0.6436 & 13.9\%  \\
12 & 3.6239 & 1.0128 & 21.8\%  \\
8 & 3.3367 & 1.3000 & 28.0\% \\
4 & 1.9621 & 2.6746 & 57.7\%  \\
2 & 0.9762 & 3.6605 & 78.9\% \\
\end{tabular}
\end{table}

As can be seen, the rolling horizon model significantly outperforms the fully static model (denoted by $\vert \mathcal{T} \vert$ in \Cref{Tab1BaseComparisonFull}) for each of the considered step sizes smaller than 96. In addition, the smaller the step size, the larger the improvement gets over the fully static model. At first glance, this seems counterintuitive, as the fully static model has access to information over all time slots at once, while in the rolling horizon setting this information is distributed over multiple iterations. Given that some of the constraints (EV and battery) connect consecutive time slots, one may expect that not having full information at once should lead to worse solutions. However, the structure of the uncertainty allows for additional uncertainty information within the individual iterations, decreasing the impact of the distributed information over the whole time horizon.\\

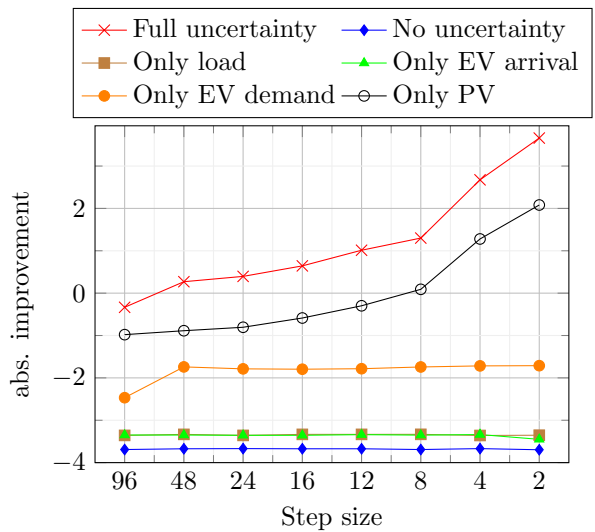
\begin{figure}
\centering
\pgfplotsset{every axis legend/.append style={at={(1.05, 1.35)}, anchor = north east, legend columns = 2}}
\begin{tikzpicture}
\begin{axis}[xmin=0.5, xmax=8.5, ymin=-4, ymax=3.95,
xlabel={Step size},
ylabel={abs. improvement},
xtick={1,2,3,4,5,6,7,8},
xticklabels={96,48,24,16,12,8,4,2},
grid = both,
minor tick num = 1,
major grid style = {lightgray},
minor grid style = {lightgray!25},
legend cell align={left},
width = 0.65*\textwidth,
height = 0.5*\textwidth]
\addplot[red, mark = x, mark size = 3]table[col sep = semicolon, y={full}]{ResultsPart1.csv};
\addplot[blue, mark = diamond*]table[col sep = semicolon, y={no}]{ResultsPart1.csv};
\addplot[brown, mark = square*]table[col sep = semicolon, y={Load}]{ResultsPart1.csv};
\addplot[green, mark = triangle*]table[col sep = semicolon, y={EV arrival}]{ResultsPart1.csv};
\addplot[orange, mark = *]table[col sep = semicolon, y={EV demand}]{ResultsPart1.csv};
\addplot[black, mark = o]table[col sep = semicolon, y={PV}]{ResultsPart1.csv};
\legend{Full uncertainty, No uncertainty, Only load, Only EV arrival, Only EV demand, Only PV}
\end{axis}
\end{tikzpicture}
\caption{Absolute improvements of different step sizes over the fully static model for different uncertainty configurations and step sizes}
\label{FigRelImpro}
\end{figure}

In the following, we discuss the impact of the individual uncertainty sources on the objective value to identify which of the underlying uncertainties are responsible for the observed improvements. In a first step, we verify the statement regarding the optimality of a rolling horizon by running the same instance, but without any uncertainty, apart from the market data. Afterward, to analyze the impact of uncertainty in more detail, we create four new scenarios, in which we always add only one uncertainty source (EV arrival, EV demand, PV, and load) to the instance without any uncertainty. In \Cref{FigRelImpro}, the absolute differences in average objective value of the five modified scenarios for the various time steps with the fully static model are given. For a better comparison, also the above-presented results for the base scenario $B$ are added to \Cref{FigRelImpro}, denoted by \textit{full uncertainty}.\\

\begin{itemize}
\item The modified scenario with \textit{no uncertainty} (blue rhombi in \Cref{FigRelImpro}) shows the previously anticipated results that the rolling horizon models with multiple iterations perform worse compared to the fully static model solving the problem once for the whole time horizon. This difference in objective value is attributed to the shorter time horizons of the iterations of the rolling horizon model. Instead of having access to all data and solving all three days at once, the individual iterations of the rolling horizon only consider 12 to 36 hours in advance. This limited information per iteration leads to suboptimal solutions. \item In case, all parameters apart from the \textit{EV arrival} (and departure) times are known beforehand, (green triangles), the fully static model performs better than any rolling horizon model (see \Cref{FigRelImpro}). Comparing the differences to the scenario without any uncertainty, some small improvements can be observed. These improvements can be explained by the updated information on EV arrival times, which may result in longer periods where an EV can be charged or discharged compared to the worst-case arrival times.
\item For the second uncertain EV parameter, the \textit{EV demand} (orange circles), the rolling horizon model performs worse than the fully static model for all step sizes. However, it performs significantly better than the previous cases of no uncertainty or uncertain EV arrivals (see \Cref{FigRelImpro}). This improvement follows from how robust optimization deals with the uncertain EV demand. As the solution has to be feasible for every possible realization of the demand, the achieved solution charges at least the maximum demand of the next trip. However, within the rolling horizon approach, at some iteration, the true EV demand is revealed. As the true realizations are drawn uniformly at random from the uncertainty sets, the realized EV demand is smaller than the maximum demand. The resulting surplus in the EV battery can now either be sold at one of the electricity markets or future charging may be reduced. This also explains the differences in improvement between a step size of 96 and the remaining (smaller) step sizes. In the case of a step size of 96, the last iteration starts at 12 pm on day 2, and therefore can not react to any realization of EV demand during the last 1.5 days. With a step size of 48 on the other hand, the last iteration starts only at 12 pm on day 3 and can therefore react to more EV demand realizations.
\item The modified uncertainty scenario with \textit{only PV} uncertainty (black empty circles) is the only scenario apart from the full uncertainty scenario, in which some of the rolling horizon models outperform the fully static model. These improvements can be explained as follows: Due to the dynamic uncertainty sets of the PV production, the models get access to additional information in the form of smaller PV uncertainty sets for some time slots during each iteration. The smaller the step size of the rolling horizon model, the more often the model gets access to this additional information, and the better the results.
\item The last uncertainty scenario considers only the household \textit{load} to be uncertain (brown squares). As can be seen in \Cref{FigRelImpro}, the rolling horizon models do not perform any better than the fully static model. This can be explained by the fact that no further information on future uncertainties become available over time. Therefore, observed realizations do not lead to additional useful information for the rolling horizon models.
\end{itemize}

When analyzing \Cref{FigRelImpro} in more detail, we notice that the shape of scenario $B$ (full uncertainty) is a composition of the scenarios with only EV demand and only PV. The first sharp improvement from step size 96 to 48 can be mainly attributed to the observations of the uncertain EV demand, while for the remaining step sizes, the shape of the curve resembles the shape of the PV scenario.\\

Summarizing, we note that mainly two uncertain parameters are responsible for the improved performance of the classical rolling horizon model, namely the EV demand and the dynamic PV uncertainty sets. In its base, the rolling horizon-based approach allows to better adapt to uncertainty realizations as well as improved forecasts. The above analysis, therefore, strengthens our choice of a rolling horizon as a suitable framework for modeling and analyzing uncertainty in energy management systems. The results also show that the step size has a significant impact on the quality of the results and that smaller step sizes generally lead to improved solutions. Nevertheless, it may not be desirable to run the model every single time slot, as computation power is still costly and energy-intense. Furthermore, each iteration requires an information exchange between households and the MGO, and not all iterations of the classical rolling horizon model may improve the solution (e.g. in the middle of the night, no improved PV forecasts nor from the observation of the actual EV demand may be expected).

\section{The Dynamic Rolling Horizon}\label{SResultsDRH}
As concluded in the previous section, there is room for improvement as not all iterations of the classical rolling horizon model offer significant further information to the model. Hence, skipping these iterations is expected to yield the same or comparable results. Based on these insights, we design a dynamic scheduling tool, which finds improved starting time slots for a rolling horizon model and we compare the resulting dynamic rolling horizon model with the classical model.

\subsection{Improved Scheduling of Rolling Horizon Iterations}\label{SSKnapsack}
Based on the insights gained in \Cref{SSBaseComparison}, the goal of the scheduling tool is to find starting time slots for the rolling horizon, which add as much information as possible. We model this tool as a version of a knapsack problem, where the items correspond to the time slots, in which an iteration of the rolling horizon model may start. In a standard knapsack, there is a 1-to-1 relation between choosing an item and gaining its value. however, this is not the case in our problem, as the information gain of choosing a time slot $s$ also depends on choices of other close-by starting time slots. To model this connection, we introduce the parameters $v_{t,s}$ and $w_{t,s}$, which represent the gained information due to the dynamic PV uncertainty sets and the EV demand in time slot $t$, given a starting time slot $s$.\\

The value $v$ represents the possible revenue due to the improvement of the dynamic PV uncertainty sets and can be attributed to the smaller uncertainty set of a time slot $t$, the closer we get to $t$. Hence, we define
\begin{equation}
v_{t,s} = p^{PV}_t \alpha^{PV}_t \beta^{PV}_{t,s} \pi^{ID,sell}_t(1-\alpha^{ID,sell}_t),
\end{equation}
where $p^{PV}_t$ and $\alpha^{PV}_t$ are the known values for the PV production of time slot $t$ (predicted at time slot 0) and $\pi^{ID,sell}_t(1-\alpha^{ID,sell}_t)$ corresponds to the worst-case selling price at the intraday market at time $t$. The value $\beta^{PV}_{t,s} \in \left[ 0,1\right] $ specifies the expected reduction factor of the uncertainty interval for time slot $t$, given an improved forecast in time slot $s$. Note that $\beta^{PV}_{t,s}$ increases, is $s$ and $t$ are closer to each other. For $s>t$, we define $\beta^{PV}_{t,s}=0$, as the realization of $s$ is already known at time slot $t$. Summarizing, $v_{t,s}$ represents the average or expected additional revenue of starting an iteration at time slot $s$, given an improved forecast for time slot $t$.\\

The information gain of the EV demand reflects that in expectation not all realizations of uncertain parameters are equal to their worst-case scenario. Hence, when an EV arrives back home, we can expect that its SoC is in general higher than expected in the worst-case. This surplus of electricity can either be sold at the intraday market, or it reduces the amount of energy that has to be bought for the next trip. This leads to 
\begin{equation}
w_{t,s} = \sum_{ev \in \mathcal{EV}(t)}P^{EV}_{ev,t}\alpha^{PV}_{t}\max_{l \geq s}\left\lbrace \pi^{ID,sell}_l(1-\alpha^{ID,sell}_l) \right\rbrace,
\end{equation}
for all $s,t \in \mathcal{T}$ with $s>t$. Hereby, $P^{EV}_{ev,t}\alpha^{PV}_{t}$ represents the expected average surplus of electricity of the EV. Note, that the set $\mathcal{EV}(t)\subseteq \mathcal{N}_{EV}$ depends on the time slot $t$ and contains all EVs, which have just arrived at time slot $t$. Hence, $w_{t,s}$represents the 'average' improvement over the worst-case EV demand only for EVs that have just arrived at $t$ and can use this surplus at the current time slot $s$ or any future time slot.\\

In the following, we introduce the decision variables of the proposed model as well as the different constraints and the objective function. Let $x_s \in \left\lbrace 0,1 \right\rbrace$, $s \in \mathcal{T}$, be a binary variable indicating whether we choose time slot $s$ as a starting time slot for an iteration of the dynamic model or not. We define the weight of all items (time slots) to be 1 and let $k$ be the upper bound on the number of iterations of the rolling horizon. This leads to the classical knapsack constraint limiting the number of chosen iterations:
\begin{equation}\label{M3KC}
\sum_{s \in \mathcal{T}}x_s \leq k.
\end{equation}
We introduce two types of auxiliary binary variables $y_{t,s}$ and $z_{t,s}$ for $t,s \in \mathcal{T}$ to link the decisions of variables $x_s$ to the information gain in time slot $t$. The following constraints model the link between choosing a time slot $s$ and its possible information gain,
\begin{equation}\label{M3LC1}
x_s \geq y_{t,s} \quad \forall s,t \in \mathcal{T},
\end{equation}
\begin{equation}\label{M3LC2}
x_s \geq z_{t,s} \quad \forall s,t \in \mathcal{T}.
\end{equation}
While the decision to choose a time slot $s$ may have an impact on multiple time slots $t$, we assume that the gains of a time slot $t$ are assigned to at most one time slot $s$ (e.g., the surplus of electricity in the EV can only be sold once). This lead to the following constraints:
\begin{equation}\label{M3KC1}
\sum_{s \in \mathcal{T}}y_{t,s} \leq 1 \quad \forall t \in \mathcal{T},
\end{equation}
\begin{equation}\label{M3KC2}
\sum_{s \in \mathcal{T}}z_{t,s} \leq 1 \quad \forall t \in \mathcal{T}.
\end{equation}
Note that per chosen time slot $s$, information of multiple different time slots $t$ can be used. The objective function now aims to maximize the sum of all weights multiplied by their respective $y$ or $z$ variables
\begin{equation}\label{M3OF}
\max \sum_{s,t \in \mathcal{T}}\left( v_{t,s}y_{t,s}+\eta w_{t,s}z_{t,s}\right),
\end{equation}
where $\eta$ represents an internal weight for the two parts of the objective function.\\

Objective function (\ref{M3OF}), together with constraints (\ref{M3KC})-(\ref{M3KC2}) form the complete model, which can be solved using standard integer linear programming solvers. Note, that the actual objective value is not used further, but only the vector $x$, which indicates the starting time slots of the rolling horizon model. In the following, we refer to the rolling horizon with starting time slots based on the knapsack problem as the \textit{dynamic rolling horizon model}.

\subsection{Comparison between classical and dynamic models}
In the following, we compare the classical and the dynamic rolling horizon models with different starting time slots with each other. Again, the data presented in \Cref{SResultsCRH} is used. We test the two models for various step sizes and the corresponding number of iterations. Note that each step size implies a fixed number of iterations throughout the time horizon of 3 days, and both parameters are used interchangeably. The considered step sizes are 96, 48, 24, 16, 12, 8, 4, and 2 with each step size being tested for all three uncertainty scenarios $A$, $B$, and $C$ (see \Cref{TabScenarios}). In addition to the average cost, we also use the average PV usage as a measure of how efficiently the improved starting times of the dynamic rolling horizon can capture the time-dependent PV uncertainty sets. This choice is based on the insights gained during the analysis of \Cref{FigRelImpro}, which indicated that the dynamic PV uncertainty sets are the driving factor behind the observed improvements of the rolling horizon models. The achieved results for the three scenarios are given in Tables \ref{Tab3Comparison1}, \ref{Tab3Comparison2} and \ref{Tab3Comparison3}.\\

\begin{table}[h]
\centering
\caption{Comparison of classical and dynamic RH, Scenario $A$}
\label{Tab3Comparison1}
\begin{tabular}{c | c c c} 
Step size & classical RH & dynamic RH & rel. improv.  \\ 
\hline
$\vert \mathcal{T} \vert$ & -10.8099 & - & -\\
96 & -10.1937 & -10.1937 & 0.0\%  \\ 
48 & -10.5697 & -10.6399 & 0.7\%  \\
24 & -10.6338 & -10.9147 & 2.6\% \\
16 & -10.7521 & -11.1579 & 3.8\%  \\ 
12 & -10.9201 & -11.3025 & 3.5\%  \\
8 & -11.0491 & -11.5721 & 4.7\%  \\
4 & -11.5565 & -11.9146 & 3.1\% \\
2 & -11.9363 & -12.1420 & 1.7\% \\
\end{tabular}
\end{table}

Starting with scenario $A$, we notice that both rolling horizon models only need a few iterations (step size 12 for the classical rolling horizon, respectively 24 for the dynamic version) to surpass the fully static robust model. Due to the small uncertainty sets within this scenario, the information gains of the rolling horizon schemes are relatively small compared to the advantage of considering the complete time horizon at once. Hence, it needs some iterations to build up the required information gain to surpass the advantage of considering the complete time horizon.\\

Focusing on the comparison between the classical and the dynamic rolling horizon models, we observe that the improvement of the dynamic model first increases with additional iterations, but then decreases again as soon as the step size gets smaller. As for small step sizes, the classical rolling horizon model already covers most of the important starting time slots over the day, and it can already make use of most of the additional information of the dynamic uncertainty sets. Note that for a step size of 1, both models start an iteration at every time slot, resulting in the same solution.\\

\begin{table}[h]
\centering
\caption{Comparison of classical and dynamic RH, Scenario $B$}
\label{Tab3Comparison2}
\begin{tabular}{c | c c c} 
Step size & classical RH & dynamic RH & rel. improv.  \\ 
\hline
$\vert \mathcal{T} \vert$ & 4.6367 & - & -\\
96 & 4.9705 & 4.9705 & 0.0\%  \\ 
48 & 4.3666 & 3.8631 & 11.5\%  \\
24 & 4.2409 & 3.2276 & 23.9\% \\
16 & 3.9931 & 2.7623 & 31.1\%  \\ 
12 & 3.6239 & 2.4049 & 33.6\%  \\
8 & 3.3367 & 1.8427 & 44.8\%  \\
4 & 1.9621 & 0.9726 & 50.4\% \\
2 & 0.9762 & 0.4192 & 57.1\% \\
\end{tabular}
\end{table}

\begin{figure}
\centering
\pgfplotsset{every axis legend/.append style={at={(0.7, 1.15)}, anchor = north east, legend columns = 3}}
\begin{tikzpicture}
\begin{axis}[xmin=-5, xmax=155, ymin=0, ymax=2,
xlabel={Number of Iterations of the RH},
ylabel={abs. improvement},
grid = both,
minor tick num = 1,
major grid style = {lightgray},
minor grid style = {lightgray!25}]
\addplot[blue, mark = x, mark size = 3]table[col sep = semicolon, y={A}]{ResultsPart2.csv};
\addplot[red, mark = square*]table[col sep = semicolon, y={B}]{ResultsPart2.csv};
\addplot[green, mark = diamond*]table[col sep = semicolon, y={C}]{ResultsPart2.csv};
\legend{$A$, $B$, $C$}
\end{axis}
\end{tikzpicture}
\caption{Absolute improvements in the objective value between classical and dynamic RH for different step sizes and uncertainty scenarios}
\label{FigAbsImprov2}
\end{figure}
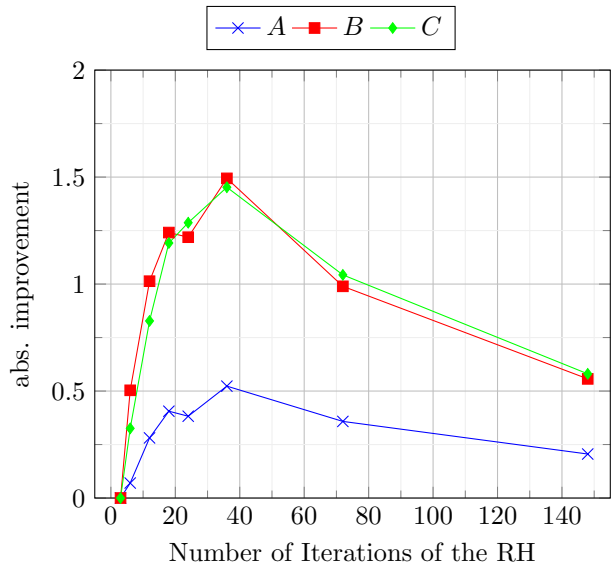

For the standard scenario $B$, both rolling horizon models outperform the fully static robust model for step sizes smaller or equal than 48 (see \Cref{Tab3Comparison2}). When comparing the two rolling horizon schemes with each other, the structure of the relative improvements differs from the previous case. Instead of having a peak at a step size of 8, followed by a decrease, the relative improvements steadily increase further. This can be explained by the fact that the values tend towards 0 and by that already small improvements can lead to large relative deviations. If instead, the absolute values, as shown in \Cref{FigAbsImprov2}, are analyzed, the same structure of an increase in improvements up to a step size of 8, followed by a decrease in improvements can be observed.\\

\begin{table}[h]
\centering
\caption{Comparison of classical and dynamic RH, Scenario $C$}
\label{Tab3Comparison3}
\begin{tabular}{c | c c c} 
Step size & classical RH & dynamic RH & rel. improv.  \\ 
\hline
$\vert \mathcal{T} \vert$ & 22.6208 & - & -\\
96 & 20.3452 & 20.3452 & 0.0\%  \\ 
48 & 19.2826 & 18.9569 & 1.7\%  \\
24 & 19.0863 & 18.2586 & 4.3\% \\
16 & 18.8243 & 17.6324 & 6.3\%  \\ 
12 & 18.4443 & 17.1576 & 7.0\%  \\
8 & 18.0163 & 16.5639 & 9.1\%  \\
4 & 16.5472 & 15.5044 & 6.3\% \\
2 & 15.4005 & 14.8201 & 3.8\% \\
\end{tabular}
\end{table}

In scenario $C$, both rolling horizon models outperform the fully static robust model for all step sizes (see \Cref{Tab3Comparison3}). The results for scenario $C$ support the previous analysis, in that up to a step size of 8, the improvements of the dynamic model steadily increase. Only for step sizes 2 and 4, the improvements are getting smaller, due to the better placement of iterations of the classical rolling horizon model.\\

\begin{figure}
\centering
\pgfplotsset{every axis legend/.append style={at={(0.7, 1.15)}, anchor = north east, legend columns = 3}}
\begin{tikzpicture}
\begin{axis}[xmin=-5, xmax=155, ymin=0, ymax=15,
xlabel={Number of Iterations of the RH},
ylabel={rel. improvement in \%},
grid = both,
minor tick num = 1,
major grid style = {lightgray},
minor grid style = {lightgray!25}]
\addplot[blue, mark = x, mark size = 3]table[col sep = semicolon, y={ArelImpro}]{ResultsPart4.csv};
\addplot[red, mark = square*]table[col sep = semicolon, y={BrelImpro}]{ResultsPart4.csv};
\addplot[green, mark = diamond*]table[col sep = semicolon, y={CrelImpro}]{ResultsPart4.csv};
\legend{$A$, $B$, $C$}
\end{axis}
\end{tikzpicture}
\caption{Relative PV usage improvements between classical and dynamic RH for different step sizes and different uncertainty scenarios}
\label{FigRelPVImpro}
\end{figure}
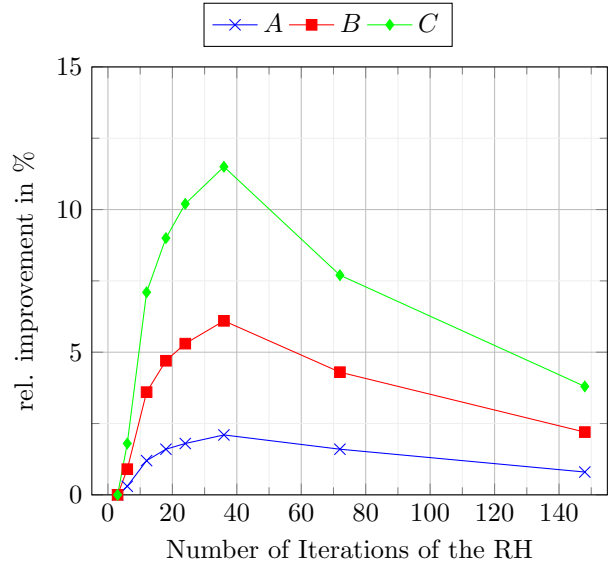

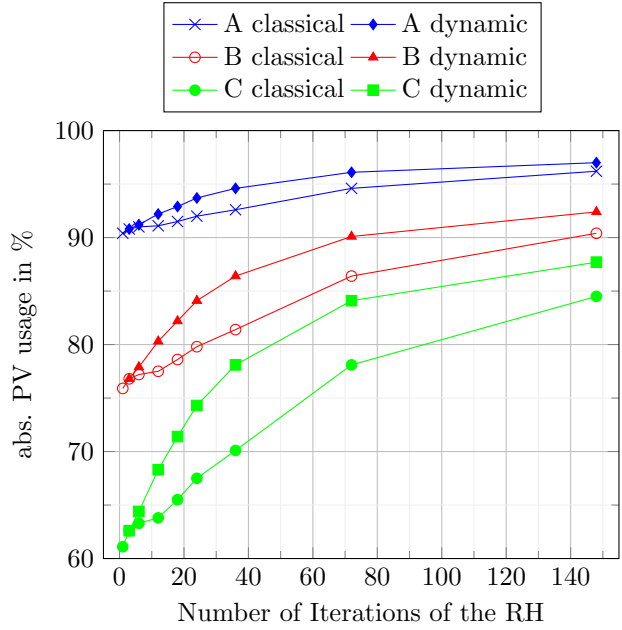
\begin{figure}
\centering
\pgfplotsset{every axis legend/.append style={at={(0.85, 1.3)}, anchor = north east, legend columns = 2}}
\begin{tikzpicture}
\begin{axis}[xmin=-5, xmax=155, ymin=60, ymax=100,
xlabel={Number of Iterations of the RH},
ylabel={abs. PV usage in \%},
grid = both,
minor tick num = 1,
major grid style = {lightgray},
minor grid style = {lightgray!25}]
\addplot[blue, mark = x, mark size = 3]table[col sep = semicolon, y={AclasRH}]{ResultsPart3.csv};
\addplot[blue, mark = diamond*]table[col sep = semicolon, y={AdynRH}]{ResultsPart3.csv};
\addplot[red, mark = o]table[col sep = semicolon, y={BclasRH}]{ResultsPart3.csv};
\addplot[red, mark = triangle*]table[col sep = semicolon, y={BdynRH}]{ResultsPart3.csv};
\addplot[green, mark = *]table[col sep = semicolon, y={CclasRH}]{ResultsPart3.csv};
\addplot[green, mark = square*]table[col sep = semicolon, y={CdynRH}]{ResultsPart3.csv};
\legend{A classical, A dynamic, B classical, B dynamic, C classical, C dynamic}
\end{axis}
\end{tikzpicture}
\caption{Absolute PV usage for different uncertainty configurations and rolling horizon models}
\label{FigPVImpro}
\end{figure}

For the local PV usage, \Cref{FigRelPVImpro} displays the relative improvements of the dynamic rolling horizon models over the classical ones, and \Cref{FigPVImpro} displays the absolute PV usage percentages. As can be seen in \Cref{FigRelPVImpro}, the dynamic rolling horizon models outperform the classical ones throughout all step sizes, and the overall shape of the relative improvements are rather similar to the (absolute) improvements in objective value, as displayed in \Cref{FigAbsImprov2}. This aligns well with our conclusion in \Cref{SResultsCRH}, which highlighted the impact of the dynamic uncertainty sets on the objective value. Furthermore, the dynamic scheduling tool, introduced in \Cref{SSKnapsack}, works as intended and significantly increases the PV usage through clever scheduling of the starting time slots of the rolling horizon (see also \Cref{FigPVImpro}). In addition, the values displayed in \Cref{FigPVImpro} also align well with the dynamic uncertainty sets of the PV generation for each of the three uncertainty scenarios $A$, $B$, and $C$. While the fully static model only uses roughly $1-\alpha^{PV}$ of the realized PV generation, which fully aligns with the robust approach, both rolling horizon models quickly improve upon this usage.\\

In general, we notice that the dynamic rolling horizon model better coordinates the iterations of the rolling horizon framework and thereby improves the handling of the considered uncertainty. The dynamic model allows to derive tailor-made starting times, which are able to capture two large sources of uncertainty, namely the uncertain PV and EV demand. As a consequence, the dynamic rolling horizon model can achieve similar or even better solutions with less than half the number of iterations (see Tables \ref{Tab3Comparison1} - \ref{Tab3Comparison3}).\\

\begin{figure}[h]
\centering
\includegraphics[width=8.5cm]{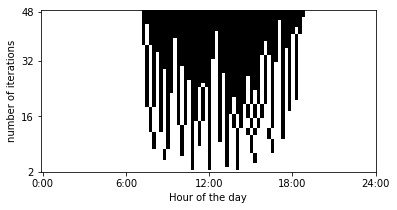}
\caption{Selected starting time slots (in black) of the dynamic scheduling tool depending on the chosen number of iterations (y-axis).}
\label{FigStarting}
\end{figure}

\Cref{FigStarting} shows the selected starting time slots resulting from the knapsack model for one day for different numbers of iterations. It can be seen that starting time slots during the day and in the early evening are preferred to time slots throughout the night when no PV generation or EV arrivals are present. In addition, it can be observed that even with very few starting time slots, the knapsack problem spreads these slots evenly over the time window of PV production. As the number of starting time slots increases, the gaps between the chosen starting time slots decrease, and more evening time slots are included. For 48 starting time slots, which correspond to a step size of 2, every time slot with PV production, as well as some early evening time slots are included. It should be noted, that the overhead due to solving the knapsack problem is negligible compared to a single iteration of the rolling horizon model. The size of the knapsack problem only depends on the time horizon $\mathcal{T}$ and is independent of the number of households or PV systems. Although the number of EVs is included in the computation of the values $w_{t,s}$, it does not change the number of variables or constraints.\\

An interesting observation beyond the comparison of the two rolling horizon models is the dependency of the objective value on different uncertainty sets for the market prices. To isolate the impact of the uncertainty within the market prices on the solution, we consider a setting, where we fix the uncertainty sets of all remaining uncertain parameters to those of scenario $B$ while using the uncertainty sets of scenarios $A$, $B$, and $C$ for the market prices. In \Cref{Tab4MarketUncertainty}, the total bought and sold electricity in kWh for each of the modified scenarios is given. One can notice, that for smaller price uncertainty sets, the MGO is more active in the electricity markets and buys and sells more electricity, although the net amount of bought electricity to cover the demand stays the same. Therefore, the additional amount of electricity is only bought and sold to increase profit. Note, that even in scenario $C$, the MGO buys electricity beyond what is needed to cover the demand. Increasing the uncertainty sets of both markets even further to $\alpha_{DA}=0.8$ and $\alpha_{ID}=0.9$ results in a solution in which no electricity is sold anymore. If this behavior of buying and selling to increase profit is undesired, one can easily change the objective function to align with other goals, such as maximizing local consumption or minimizing the amount of bought electricity.

\begin{table}[ht]
\centering
\caption{Impact of Market Uncertainty on the Market Decisions}
\label{Tab4MarketUncertainty}
\begin{tabular}{c | c c c} 
Scenario: & $A^*$ & $B$ & $C^*$  \\ 
\hline
bought energy & 1,848.54 & 1,450.49 & 1,219.15\\
sold energy & 916.42 & 518.25 & 284.59  \\ 
net bought energy & 932.12 & 932.24 & 934.56 \\
\end{tabular}
\end{table}

\section{Conclusion}\label{SConclusion}
In this paper, we presented a robust energy management approach for microgrids. Furthermore, we merged this robust model with a rolling horizon framework to allow for a better representation of the uncertainty within the decision-making process, as well as to be able to react to the realizations of uncertain parameters. This rolling horizon-based model shows promising first results and a detailed analysis highlighted the reasons behind the improved solutions. Based on these insights, a dynamic scheduling tool, which further improves the solutions by up to 57\% in terms of costs, was introduced. This tool is based on structural knowledge of the uncertainty sets to find good starting time slots for the rolling horizon framework and results in overall better solutions given the same or smaller number of iterations of the rolling horizon. \\

Given these first promising insights into the dynamic scheduling of a rolling horizon framework in the context of local energy management, some interesting future research directions arise. In a first step, the integration of probability distributions instead of the simple usage of a uniform distribution may lead to further insights and improvements. Another interesting direction is to transform the offline computation of the starting time slots into an online version. Instead of determining the set of starting time slots for the rolling horizon in advance for the overall time horizon, it may be beneficial to decide in an online fashion, whether to start an iteration or not. One important advantage of such an approach is that the realizations of some uncertain parameters are already known and that an online algorithm can therefore better react to these realizations. Hence, online algorithms have access to more information than the presented offline version, leading to hopefully even better starting time slots.

\section*{Acknowledgements}
This research is supported by the Netherlands Organization for Scientific Research (NWO) Grant 645.002.001.

\printbibliography

\end{document}